\documentclass{ws-p8-50x6-00}
\newcommand{\PO}{\rm l \! P }
\newcommand{\RO}{\rm l \! R }
\newcommand{\xpom}{x_{\PO} }

\begin{document}

\title{High energy QCD and hard diffraction at HERA versus Tevatron}

\author{C. Royon}

\address{Service de Physique des Particules,
CE-Saclay, F-91191 Gif-sur-Yvette Cedex, France;
Brookhaven National Laboratory, Upton, New York, 11973;
University of Texas, Arlington, Texas, 76019}


\maketitle

\abstracts{We present and discuss two different topics where Tevatron
and HERA data can be compared directly.
A method allowing for a direct comparison of data with theoretical  predictions 
is proposed for forward jet production at HERA. An application to the determination 
of the  effective Pomeron intercept in 
the BFKL-LO parametrization from $d\sigma/dx$ data at HERA   leads to a good fit 
with a significantly high effective 
intercept, $\alpha_P= 1.43 \pm 0.025 (stat.) \pm 0.025 (syst.)$. It is 
less than the value of the pomeron
intercept using dijets with large rapidity intervals obtained at Tevatron.
In a second part of this report, we make comparison between diffractive
results at HERA and at Tevatron. We first give the parton distributions
in the pomeron extracted from HERA data, and compare with hard single
diffraction at Tevatron and diffractive dijets with a leading antiproton data.
}

\section{Forward jets at HERA and Mueller-Navelet jets at Tevatron}

\subsection{Forward jets at HERA}

The study of forward jets at $ep$ colliders is considered as the milestone of QCD 
studies at high energies,
since it provides a direct way of testing the perturbative resummations of soft 
gluon radiation. It is similar to the previous proposal of studying two jets 
separated by a large rapidity interval in hadronic 
colliders~\cite{mu}, for which preliminary results are 
available~\cite{D0}.

The cross-section for forward jet production at HERA in the dipole model 
reads~\cite{ba92}:
\begin{eqnarray}
&~& \frac{d^{(4)} \sigma}{dx dQ^2 dx_J d k_T^2 d \Phi} =
\frac{ \pi N_C \alpha^2 \alpha_S(k_T^2)}{Q^4 k_T^2} 
\ f_{eff} (x,\mu_f^2)
\ \Sigma e_Q^2 
\int_{\frac 12- i\infty}^{\frac 12+ i\infty} \frac{d \gamma}{2i \pi} 
\left( \frac{Q^2}{k_T^2} \right)^{\gamma} \times 
\nonumber \\
&~& \times \exp \{\epsilon (\gamma,0) Y\} \left[ \frac{h_T(\gamma) +h_L(\gamma)}{\gamma}
(1-y) + \frac{h_T(\gamma)}{\gamma} 
\frac{y^2}{2} \right]
\label{dsigma}
\end{eqnarray}
where
\begin{eqnarray}
Y &=& \ln \frac{x_J}{x} \\
\epsilon(\gamma,p)&=& \bar{\alpha} \left[ 2 \psi(1) -\psi(p+1-\gamma)
-\psi (p+\gamma) \right] \\
f_{eff} (x, \mu_f^2) &=& G(x,\mu_f^2) + \frac{4}{9} \Sigma (Q_f+ \bar{Q_f}) \\
\mu_f^2 &\sim& k_T^2\ ,
\end{eqnarray}
are, respectively, $Y$ the rapidity interval between the photon probe and the 
jet,
$\epsilon(\gamma,p)$ the BFKL kernel eigenvalues, $f_{eff}$ the effective 
structure 
function 
combination, and $\mu_f$ the corresponding factorization scale.  
The main BFKL parameter is $\bar{\alpha},$ which is the (fixed) value of the 
effective 
strong coupling constant in LO-BFKL formulae. 

The so-called ``impact factors'' $h_T$ and $h_L$
are obtained from the $k_T$ factorization properties~\cite{ca91} of the coupling 
of 
the BFKL amplitudes to external hard probes. The same factors can be related 
to the photon wave functions within the equivalent context of 
the QCD 
dipole model~\cite{mu94,dipole}.

The main problem to solve is to investigate the effect of the experimental cuts on the 
determination of the 
integration variables leading to a prediction for $d\sigma/dx$ from the given 
theoretical formula for 
$d^{(4)} \sigma$ as given in formula (\ref{dsigma}). The effect appears as 
bin-per-bin {\it 
correction factors} to be multiplied to the theoretical cross-sections for 
average values of the 
kinematic variables for a given $x$-bin before comparing to data \cite{papjets}.

The experimental correction factors have been determined using
a toy Monte-Carlo designed as follows. We generate flat distributions in the 
variables $k_T^2/Q^2$, $1/ Q^2$, $x_J,$ using reference intervals   
which include the whole of the experimental phase-space (we chose the variables
which minimize the variation of the cross-section over the measured kinematical
range. The correction factors are given in Reference \cite{papjets}.
We perform a 
fit to the H1 and ZEUS data with only two free parameters. these are the {\it 
effective} strong coupling
constant in LO BFKL formulae $\bar{\alpha}$ corresponding to the {\it effective} 
Lipatov intercept
$\alpha_P= 1+4 \log 2 \bar{\alpha} N_C/\pi$, and the cross-section 
normalisation. The 
obtained values
of the parameters and the $\chi^2$ of the fit are given in Table {\bf I} for a 
fit to 
the H1 and ZEUS 
data
separately, and then to the H1 + ZEUS data together.

\begin{center}
\begin{tabular}{|c|c|c|c|c|c|} \hline
 fit & $\bar{\alpha}$ &  Norm. & $\chi^2 (/dof)$ \\ 
\hline\hline
 H1 & 0.17 $\pm$ 0.02 $\pm$ 0.01  & 29.4 $\pm$ 4.8
 $\pm$ 
5.2 & 5.7 (/9)\\
 ZEUS & 0.20 $\pm$ 0.02 $\pm$ 0.01  & 26.4 $\pm$
 3.9 $\pm$ 
4.7 & 2.0 (/2)\\
 H1+ZEUS & 0.16 $\pm$ 0.01 $\pm$ 0.01  & 30.7
 $\pm$ 2.9 
$\pm$ 3.5 & 12.0 (/13)\\
\hline
\end{tabular}
\end{center}
\vskip .5cm
\begin{center}
{Table I- Fit results}
\end{center}

\begin{figure}
\begin{center}
\centerline{\psfig{figure=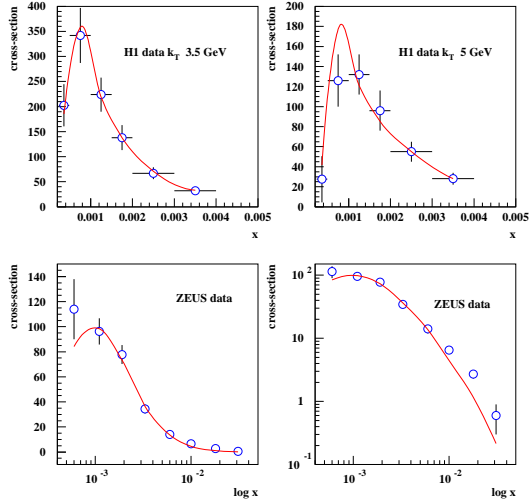,height=3.in}}
\end{center}
\caption{The H1 data ($k_T > 3.5$ GeV, $k_T > 5$ GeV), and the ZEUS data are
compared with the result of the fit. ZEUS data are also displayed in 
logarithmic scales in vertical coordinates to show the discrepancy at high
$x$ values.}
\end{figure}

\subsection{Mueller Navelet Jets at Tevatron (D0)}

It is fruitful to compare our results with the effective intercept 
we obtain from recent preliminary dijet data obtained by the
D0 Collaboration at Tevatron \cite{D0}. The measurement consists
in the ratio $R=\sigma_{1800}/ \sigma_{630}$ where $\sigma$ is the dijet 
cross-section at large rapidity interval $Y \sim \Delta \eta$ 
for two center-of-mass energies
(630 and 1800 GeV), $\Delta \eta_{1800}=4.6$, $\Delta \eta_{630}=2.4.$
The experimental measurement is $R=2.9 \pm 0.3$ (stat.) $\pm 0.3$ (syst.).
Using the Mueller-Navelet formula \cite{mu}, this measurement allows us to get 
a value 
of the effective intercept for this
process.
We get $\alpha_P$=1.65 $\pm$ 0.05 (stat.) $\pm$ 0.05 (syst.), in agreement with
the value obtained by D0 using a saddle-point approximation \cite{D0}.
This intercept is higher than the one obtained in the forward jet study.
The question arises
to interpret the different values of the  effective intercept. It could 
reasonably
come from the differences in higher 
order QCD corrections for the BFKL kernel and/or in the impact factors
depending on the initial probes \cite{papjets}.

\section{Diffraction at HERA versus Tevatron}
\subsection{Diffraction at HERA}
The data taken in 1992 by the H1 and ZEUS experiments showed some new
interesting events with an interval of rapidity around the incident
proton direction devoid of any hadronic activity. This means that a
colourless exchange ("pomeron") 
must have occured since there is no colour 
connection between the remnant proton and the struck quark.
The selection used to tag these events is mainly by asking a gap in rapidity
in the forward proton direction \footnote{There are other selection used
by H1 and ZEUS collaborations, namely tagging a proton in the final state
or using the $M_X$ subtraction method \cite{f2d94}.}. 
These events represent about 10\% of the
total deep inelastic scattering events.

The statistics obtained in 1994 allowed a measurement
of the proton diffractive structure function defined in analogy to the
standard proton structure function in a wide kinematical domain \cite{f2d94}. The data
accumulated between 1995 and 1997 allowed to extend the measurement to lower
and higher $Q^2$. In addition to the usual deep inelastic variable $x$
(the momentum fraction of the interacting quark), and $Q^2$ (the transfered energy squared between
the electron and the interacting quark), two other kinematical variables are
used, namely $\beta$,
the momentum fraction of the colourless exchanged object, and 
$\xpom =x / \beta$, 
the momentum fraction of the parton inside this object if we assume it has
a partonic structure.

The diffractive
structure function $F_2^{D(3)}$ can be investigated in the framework of Regge
phenomenology when both a 
leading ($\PO$) and a sub-leading ($\RO$) trajectory  are considered, such
that

\begin{eqnarray}
F_2^{D(3)}(Q^2,\beta,x_{\PO})=
f_{\PO / p} (x_{\PO}) F_2^{\PO} (Q^2,\beta)
+ f_{\RO / p} (x_{\PO}) F_2^{\RO} (Q^2,\beta) \ .
\label{reggeform}
\end{eqnarray}

In this parameterisation,
$F_2^{\PO}$ can be interpreted as the structure function of the 
pomeron. The values of
$F_2^{\RO} (Q^2,\beta)$ are taken from a parameterisation of the 
pion structure function \cite{GRVpion}, with a single free normalisation. 
The pomeron flux is assumed to follow a Regge behaviour with a linear
trajectory $\alpha_{\PO}(t)=\alpha_{\PO}(0)+\alpha^{'}_{\PO} t$.

Using H1 1994 data \cite{f2d94}, the resulting 
value of $\alpha_{\PO}(0)$ is $\alpha_{\PO}(0) = 1.20 \pm 0.09$
and is significantly larger than values extracted from
soft hadronic data ($\alpha_{\PO} \sim 1.08$).
Also, we find $\alpha_{\RO}(0)=0.62 \pm 0.03$. Using ZEUS data
\cite{f2d94}, we find a pomeron intercept 
$\alpha_{\PO} = 1.127 \pm 0.040$, lower than
the H1 value \footnote{The selection
of the diffractive sample in ZEUS is different from 
H1 and uses the so called $M_X$ method \cite{f2d94}.}.

The $Q^2$ evolution of the pomeron structure function may be understood in terms
of parton dynamics and therefore perturbative QCD
where parton densities are evolved according to DGLAP  \cite{al77} equations. 
We assign parton distribution functions to the pomeron and to
the reggeon. A simple
prescription is adopted in which the parton distributions of 
both the pomeron
and the reggeon are parameterised in terms of non-perturbative input
distributions at some low scale $Q_0^2= 3$ GeV$^2$ \cite{ourpapf2d}. 

\begin{figure}
\begin{center}
\centerline{\psfig{figure=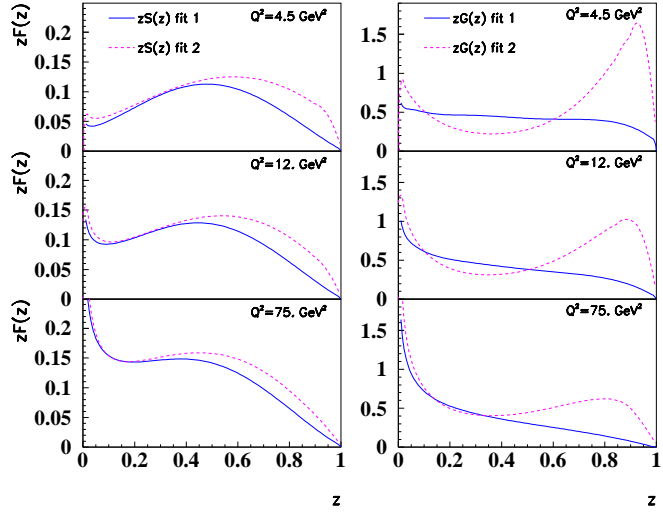,height=3.in}}
\end{center}
\caption{Quark flavour singlet ($zS$, left) and gluon ($zG$, right) distributions
of the pomeron deduced as a function of $z$, the fractional momentum of the
pomeron carried by the struck parton. 
The fit result is denoted here as fit 1. Another favourable fit is also
given.
($\chi^2/ndf = 177.1/154 = 1.15$ for fit 1
and $\chi^2/ndf = 192.5/154 = 1.25$ for fit 2 with 
statistical errors only).}
\label{f1}
\end{figure}

The resulting parton densities of the pomeron are
presented in figure \ref{f1} as a function of $z$, the fractional momentum
of the pomeron carried by the struck parton. 
We find one possible fit quoted here as fit 1. 
Fit 1
shows a large gluonic content. The quark contribution is much smaller 
compared to the gluon one. We also note that we find an other much
less favoured fit quoted here as fit 2
with a peaked gluon at high $\beta$ \cite{ourapf2d}.

We have redone this QCD analysis with ZEUS 1994 
diffractive cross-section measurements \cite{f2d94}
applying the same cuts. The gluon density is found to be lower by about a
factor 2, but the error bar on the ZEUS gluon density is large (about 50\%)
as only 30 data points are included in the fit. The quark component is found to
be similar. This difference in the gluon density can however
lead to differences in the
charm structure function as it is very much sensible to the gluon
structure function \cite{ourpapf2d}.

\subsection{Diffraction at Tevatron}

Diffractive events selection at Tevatron is basically the same as at HERA. A
rapidity gap is asked, either between the proton (antiproton) direction and
the jets inside the detector, or a rapidity central gap is asked between two
jets inside the main part of the detector.

\vspace{0.3cm}
\noindent
{\bf Hard single diffraction} \\
D0 and CDF collaborations \cite{hardsingle} have studied the fraction of
central and forward jet events with rapidity gap at two different
center-of-mass energies of 630 and 1800 GeV. The D0 results  \footnote{ CDF
results \cite{hardsingle} are compatible with the D0 ones.}
are given in Table II
and are compared to MC simulations with different pomeron structure
functions (hard gluon, $f(\beta)= \beta (1-\beta)$,
soft gluon, $f(\beta)= (1- \beta)^5$, and quarks).
It can be first noticed that the gap fractions at 630 GeV are larger than the
ones at 1800 GeV, and are much smaller than at HERA (about 1\% to be compared
with about 10\% at HERA), which can be explained because there is no
factorisation between both experiments \cite{collins}. 
The flat and hard gluon predictions are too high
compared to the data, whereas the quark scenario could work (it has however
been shown that this would lead to predict an excesive rate of diffractive W
production). A combination of soft and hard gluon could be possible to describe
this measurement, whereas the HERA data have the tendency to favour the hard
gluon scenario. The HERA and Tevatron kinematical domains in $\beta$ and 
$W$ are however different, and a more precise study including the QCD evolution
of parton distributions measured at HERA in the Tevatron domain would be of
great interest. 

\begin{center}
\begin{tabular}{|c|c|c|c|c|} \hline
 Sample & Data & Hard Gluon & Flat Gluon &  Quark \\
\hline\hline
 $1800$\,GeV  $|\eta|>1.6$ & $(0.65 \pm  0.04)\% $  &
     $(2.2 \pm 0.3)\%$ & $(2.2 \pm 0.3)\%$ &
      $(0.79 \pm 0.12)\%$ \\
 $1800$\,GeV  $|\eta|<1.0$ & $(0.22 \pm 0.05\%$   &
     $(2.5 \pm 0.4)\%$ & $(3.5 \pm 0.5)\%$ &
      $(0.49 \pm 0.06)\%$ \\
 $630$\,GeV  $|\eta|>1.6$  & $(1.19 \pm 0.08)\%$   &
     $(3.9  \pm 0.9)\%$ & $(3.1 \pm 0.8)\%$ &
      $(2.2 \pm 0.5)\%$ \\
 $630$\,GeV  $|\eta|<1.0$  & $(0.90 \pm 0.06)\%$   &
     $(5.2 \pm 0.7)\%$ & $(6.3 \pm 0.9)\%$ &
      $(1.6 \pm 0.2)\%$ \\ \hline
\end{tabular}
\end{center}
\vskip .5cm
\begin{center}
{Table II- Measured and predicted gap fractions and their ratios (D0).}
\end{center}

\vspace{0.3cm}
\noindent
{ \bf Hard color singlet exchange} \\
At Tevatron (like at HERA), it is also possible to study events with a central
rapidity gap between jets. This class of events could come from pomeron
exchange at large momentum transfer $t >> 100$ GeV$^2$, compared to the 
previous measurement which was at low $t$ ($t \sim$0). The fraction of
dijet events with a central rapidity gap is again about 1\% \cite{singlet},
to be compared with 10\% at HERA. An increase of gap fraction is observed
with jet transverse energy and rapidity.

\vspace{0.3cm}
\noindent
{\bf Diffractive dijets with a leading antiproton} \\
The CDF collaboration installed roman-pot detectors allowing to tag
$\bar{p}$ in the final state. They select a diffractive dijet subsample
of it requiring two jets of an energy greater than 7 GeV \cite{cdftag}.  

\begin{figure}
\begin{center}
\centerline{\psfig{figure=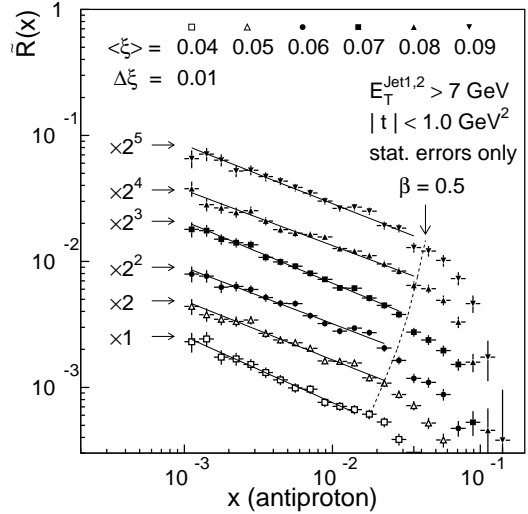,height=3.in}}
\end{center}
\caption{Ratio of diffractive to non-diffractive dijet events rates
as a function of the momentum fraction of $p$ carried by the tagged 
antiproton in the final state.}
\label{ratiocdf}
\end{figure}

Figure \ref{ratiocdf} shows the ratio of single diffractive events over the non
diffractive ones, where the two data samples are normalised to the same
luminosity, in six $\xpom$ bins (at Tevatron, $\xpom$ is called $\xi$). This
ratio does not show any $\xpom$ dependence and the $\xpom$ dependence can be
factorised out from the $x$ one, which is similar to what has been found at 
HERA except at high $\xpom$ and low $\beta$ where secondary reggeons are 
necessary to fit H1 data.

\begin{figure}
\begin{center}
\centerline{\psfig{figure=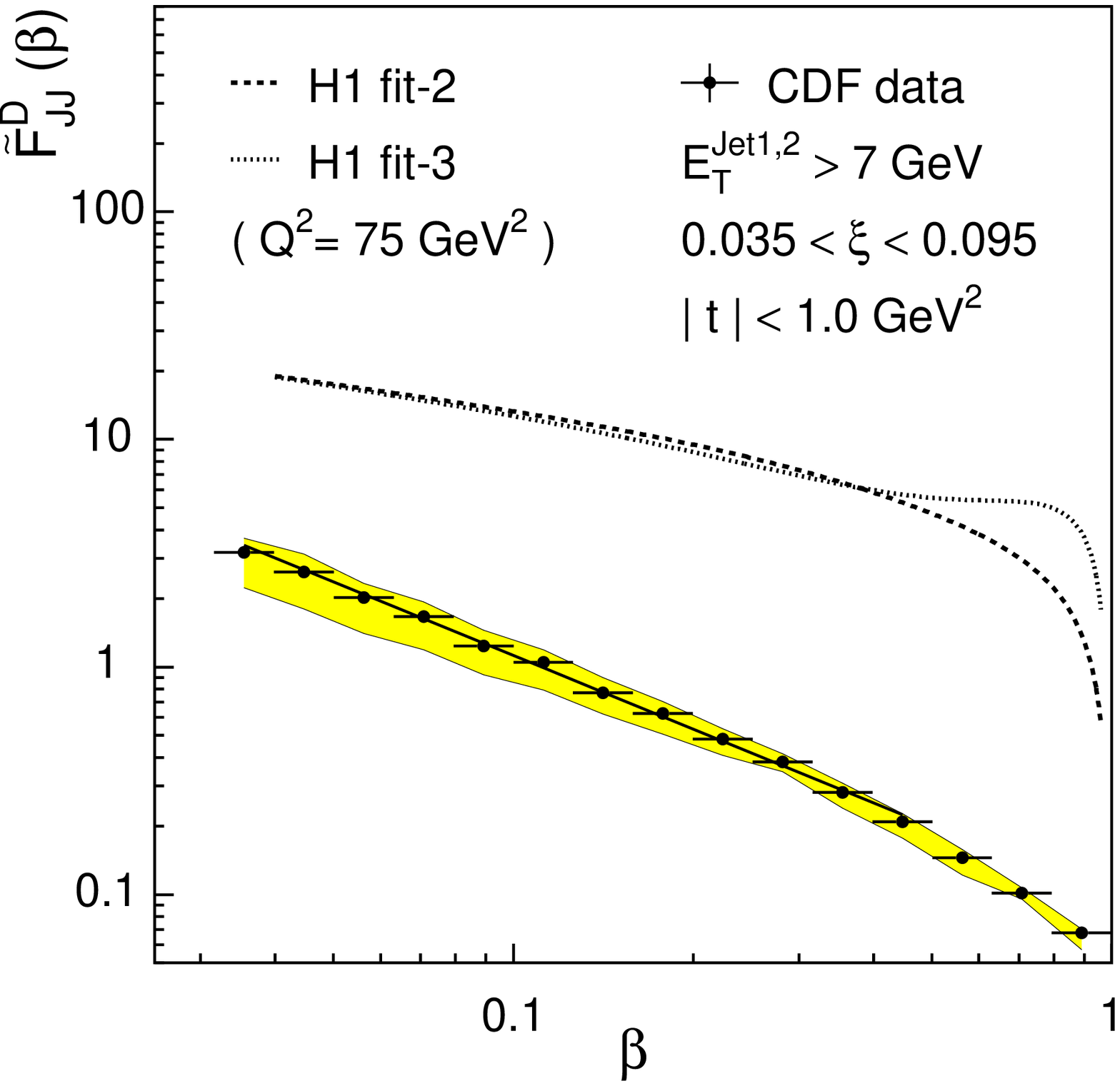,height=3.in}}
\end{center}
\caption{$\beta$ distributions of CDF data compared with predictions coming
from HERA fits. (fits 1 and 2 are designed respectively in this figure
by H1-fit 2 and H1-fit 3).}
\label{betacdf}
\end{figure}

Knowing the ratio of diffarctive to non-diffractive events, it is quite easy
to extract the pomeron structure function at Tevatron, by multiplying
this ratio by the proton structure function. For this sake, the CDF
collaboration chose to use the GRV parametrisation \cite{cdftag}.
The result together with extrapolations of H1 fits is given in Figure
\ref{betacdf}. The results disagree both in normalisation and shapes.
This disagreement represents a breaking of factorisation as expected
\cite{collins}. It is quite challenging to analyze this breaking of
factorisation which might be $\beta$ dependent using run I and run II data.
In run II, the D0 collaboration will have roman pots in both sides of
the detector allowing to tag both proton and antiproton in the final state,
it will be quite interesting to be able to test factorisation directly.

\section{Conclusion}
We have presented and discussed two related topics at HERA and Tevatron,
namely forward jets and diffraction. 

Using a new method to disantangle the effects of the
kinematic cuts from the genuine dynamical values of the forward jet 
cross-sections
at HERA, we find that the effective pomeron intercept is $\alpha_P=1.43 \pm 
0.025$
(stat.) $\pm 0.025$ (syst.). It is much higher than the soft pomeron intercept,
and, among those determined in hard processes,  it is intermediate 
between $\gamma^* \gamma^*$
interactions at LEP \cite{ro99} and dijet productions with large rapidity intervals at 
Tevatron. 

Diffraction at HERA and Tevatron give quite different results. The D0 and CDF
collaborations measure about 1\% diffractive events whereas at HERA, it is
close to 10\%. QCD analysis of HERA data lead to a pomeron made of gluons, and
a hard structure function is favored whereas the D0 collaboration showed 
that a combination of hard and soft structure functions is needed. The CDF
collaboration was able to make for the first time a direct measurement 
of the antiproton diffractive structure function, allowing a direct comparison
with HERA, by using their roman pot data. The data show large discrepancies
both in shape and in normalisation. A QCD analysis of run I Tevatron data
and run II data where D0 will have roman pot detectors on each side
are quite challenging.

\section*{Acknowledgments}
I thank J.Bartels, G.Contreras, H.Jung, R.Peschanski and L.Schoeffel for 
collaboration.

\end{document}